\documentclass[conference,compsoc]{IEEEtran}
\IEEEoverridecommandlockouts
\usepackage{cite}
\usepackage{amsmath,amssymb,amsfonts}
\usepackage{algorithmic}
\usepackage{xcolor}
\definecolor{lightgray}{RGB}{230,230,230}
\usepackage{graphicx}
\usepackage{textcomp}
\usepackage{svg}
\usepackage{listings}
\usepackage{float}
\usepackage{doi}
\usepackage{url}
\usepackage{ragged2e}
\usepackage{rotating}
\usepackage{caption}
\usepackage{array}
\usepackage[numbers]{natbib}
\usepackage{xcolor}
\usepackage{longtable}
\usepackage{pgfplots}
\usepackage{multirow}
\usepackage{tcolorbox}
\usepackage{lipsum}
\usepackage{pgfplotstable}
\usepackage{booktabs} 
\usepackage{tabularx} 
\usepackage{tikz}

\usetikzlibrary{shapes.geometric}
\pgfplotsset{compat=1.18}
\def\BibTeX{{\rm B\kern-.05em{\sc i\kern-.025em b}\kern-.08em
    T\kern-.1667em\lower.7ex\hbox{E}\kern-.125emX}}
\begin{document}

\title{Beyond Static Tools: Evaluating Large Language Models for Cryptographic Misuse Detection\\
}

{
\author{\IEEEauthorblockN{Zohaib Masood}
\IEEEauthorblockA{ \textit{Ontario Tech University}\\
Oshawa, Canada \\
zohaib.masood@ontariotechu.net}
\and
\IEEEauthorblockN{Miguel Vargas Martin}
\IEEEauthorblockA{\textit{Ontario Tech University}\\
Oshawa, Canada \\
miguel.martin@ontariotechu.ca}
}
}

\maketitle

\begin{abstract}
The use of Large Language Models (LLMs) in software development is rapidly growing, with developers increasingly relying on these models for coding assistance, including security-critical tasks. Our work presents a comprehensive comparison between traditional static analysis tools for cryptographic API misuse detection—CryptoGuard, CogniCrypt, and Snyk Code—and the LLMs—GPT and Gemini. Using benchmark datasets (OWASP, CryptoAPI, and MASC), we evaluate the effectiveness of each tool in identifying cryptographic misuses. Our findings show that GPT 4-o-mini surpasses current state-of-the-art static analysis tools on the CryptoAPI and MASC datasets, though it lags on the OWASP dataset. Additionally, we assess the quality of LLM responses to determine which models provide actionable and accurate advice, giving developers insights into their practical utility for secure coding. This study highlights the comparative strengths and limitations of static analysis versus LLM-driven approaches, offering valuable insights into the evolving role of AI in advancing software security practices.
\end{abstract}

\section{Introduction}
Protecting sensitive data on digital devices from eavesdropping or forgery relies primarily on cryptography. To ensure effectiveness in protection, the cryptographic algorithms employed must be conceptually secure, implemented accurately, and utilized securely within the relevant application. Despite the existence of mature and still secure cryptographic algorithms, numerous studies have pointed out that application developers face challenges in utilizing the Application Programming Interfaces (APIs) of libraries that incorporate these algorithms. As an illustration, Lazar et al.~\cite{b1} examined 269 vulnerabilities related to cryptography and discovered that merely 17\% are associated with flawed algorithm implementations, while the remaining 83\% stem from the misuse of cryptographic APIs by application developers. Additional research indicates that around 90\% of applications utilizing cryptographic APIs include at least one instance of misuse~\cite{b2,b3}.

Software designers and developers need to tackle this security flaw by ensuring that their applications encrypt the sensitive data they handle and store. Despite the availability of educational materials~\cite{b4} aimed at raising awareness and providing guidance on secure code development, many developers remain unaware of security considerations~\cite{b5}. Training initiatives are not universally received by programmers~\cite{b5}, and security is frequently treated as a secondary, desired goal~\cite{b6}, rather than a mandatory one, depending on the perceived risk and criticality of the developed application. Typically, developers prioritize meeting functionality and time-to-market requirements as their primary goals.

In exploring the factors contributing to this prevalent misuse, researchers previously triangulated findings from four empirical studies, including a survey involving Java developers with prior experience using cryptographic APIs~\cite{b7}. Their findings reveal that a significant majority of participants encountered difficulties in utilizing the respective APIs. Several other tools~\cite{b8,b9,b10,b11} have been developed to detect cryptographic misuses, however, the robustness of these tools have been still in question~\cite{b12,b13}. In addition to it, novice developers have started adopting AI-assisted tools to code their programming problems. Recent advancements encompass Github’s Copilot~\cite{b14}, DeepMind's AlphaCode~\cite{b15}, Amazon's Q Developer~\cite{b16}, Tabnine~\cite{b17}, Google’s Gemini~\cite{b32} and Open AI’s ChatGPT~\cite{b18}—six systems capable of translating a problem description into code. Prior results suggest that Open AI’s ChatGPT reliably produces Java programming solutions known for their elevated readability and well-organized structure~\cite{b19}.

Prior research has primarily tested Large Language Models (LLMs) using a limited set of benchmark datasets, providing an initial understanding of their performance. However, a comprehensive study evaluating the effectiveness of LLMs across a wider range of benchmarks is still lacking. Furthermore, no previous work has closely examined the quality of LLM responses, particularly with regard to the accuracy and actionability that would validate these responses for practical use. This study addresses these gaps by conducting an analysis of LLM performance across diverse datasets and introducing a framework for assessing response quality. In our work, we aim to assess the efficacy of identifying cryptographic misuses in the provided code and compare it with state-of-the-art (SOTA) static cryptographic analysis tools from literature such as CogniCrypt~\cite{b9}, CryptoGuard~\cite{b8}, and an industry-based tool, Snyk Code~\cite{b22}. To evaluate LLMs accuracy against these tools, we utilized well-known benchmark datasets employed in previous literature for cryptographic misuse detection, namely OWASP Bench~\cite{b21}, CryptoAPI Bench~\cite{b20}. In addition, we investigated LLMs robustness in detecting mutated test cases, an area where many static tools often falter~\cite{b13}. Furthermore, we evaluated the quality of responses provided by LLMs using two metrics, Actionability and Specificity, which assess whether the response can assist developers in identifying the cause of misuse and in implementing a fix.

Our research aims to address the following research questions (RQs):\\

\textit{RQ1. How effective are LLMs in detecting cryptographic misuses compared to other static tools?}

The static tools tested focus on slightly different pattern sets, leading to varied trade-offs in precision and recall, both among the tools themselves and in comparison to LLMs. To evaluate the effectiveness of LLMs in detecting cryptographic misuses compared to static tools, we ran test cases from the CryptoAPI and OWASP Benchmarks. Based on the findings, ChatGPT had a better detection rate across both benchmarks. For CryptoAPI, ChatGPT missed only 3 true instances, with CryptoGuard lagging behind at 24 misses. In the OWASP Benchmark, ChatGPT identified all true instances, while CryptoGuard missed 40 true misuse cases. However, for the OWASP Benchmark, ChatGPT had a high false positive rate, indicating that no tool is universally superior across both benchmarks.\\

\textit{RQ2. How robust are LLMs in detecting mutated test cases that other static tools fail to detect?}

To evaluate the robustness of LLMs, we ran test cases from a manually curated MASC dataset~\cite{b13} against LLMs to see if LLMs can detect these mutations of cryptographic code effectively. Our results suggest that GPT performed better than Gemini in detecting cryptographic mutations.\\

\textit{RQ3. How do prevalent LLMs compare in detecting cryptographic misuse and providing actionable, specific guidance for developers?}

To address this, we compared the performance of LLMs across both benchmarks and a mutated dataset to evaluate which LLM performs best. Additionally, we introduced a method to assess whether an LLM’s response can assist developers in fixing misuses. Since LLMs often generate text-heavy responses that may not effectively help developers, we implemented a keyword-based approach to analyze LLM outputs. This approach identifies which LLM provides more actionable and specific guidance for developers to address misuse instances.\\

Our main contributions from this work are as follows:
\begin{itemize}
    \item We are the first to conduct a comprehensive evaluation of the effectiveness of LLMs in detecting cryptographic misuses, comparing their performance to that of SOTA static analysis tools. It highlights the strengths and weaknesses of LLMs compared to static tools, offering insights into their reliability and applicability for secure software development.
    \item We assess the robustness of LLMs by evaluating their ability to detect cryptographic misuse in mutated test cases that static tools often miss. This contribution explores whether LLMs can adapt to variations in misuse patterns, providing an understanding of their resilience and potential superiority in handling unconventional or complex misuse scenarios.
    \item To systematically assess the practicality of LLM-generated guidance, we develop a keyword-based framework that scans LLM responses for actionable elements. This framework serves as a tool to measure the usability of LLM outputs for developers, supporting the identification of models that best fulfill developers' needs in addressing cryptographic misuse. To the best of our knowledge, this is the first work that explores the actionability and specificity of LLM responses.
\end{itemize}

In the next section, we cover the motivation for our work. Section III discusses background and related work. Section IV explains the methodology, while Section V presents the results and findings. Section VI covers the discussion, followed by Section VII on ecological validity. Section VIII mentions the limitations and future directions, and finally we conclude our paper in Section IX.

\section{Motivation}
To proactively identify vulnerabilities before software release, individuals without specialized expertise—such as software developers or quality assurance teams—are likely to integrate cryptographic API misuse detectors, commonly known as crypto-detectors, into the Continuous Integration/Continuous Delivery (CI/CD) pipeline. Examples of these detectors include CryptoGuard [8] and CogniCrypt~\cite{b9}. Consequently, the incapability of a crypto detector to identify an instance of misuse that it claims to detect has a direct impact on the security of end-user software. We illustrate this challenge with an example.

Imagine Johnny, a Java developer utilizing CryptoGuard [8], an advanced crypto detector, to pinpoint cryptographic vulnerabilities in her software before its release. In one of her applications, Johnny opts to employ the DES cipher, outlined as follows:\\

\textit{Cipher cipher = Cipher . getInstance (``des'');}\\

This signifies an instance of misuse involving the use of the vulnerable DES cipher. CryptoGuard falls short in recognizing this vulnerability because Johnny utilizes ``des" instead of ``DES" as the parameter~\cite{b13}. However, this presents a concern because the difference in lower case for the parameter lacks functional significance, as Java officially supports both upper case and lower case. Due to CryptoGuard's oversight of this vulnerability, Johnny might erroneously believe that his application is secure and may proceed with releasing it to end-users.

The same example was tested with ChatGPT by providing a starting prompt ``Can you detect cryptographic API misuse in the given Java code?''. ChatGPT generated the following response:

\textit{``The line of code you provided initializes a Cipher object in Java using the DES (Data Encryption Standard) algorithm. However, it's important to note that DES is considered insecure for many applications due to its small key size (56 bits) and vulnerability to brute-force attacks. It is generally recommended to use more secure algorithms like AES (Advanced Encryption Standard) instead.''}

In other words, these static tools for cryptographic misuse detection could have fundamental flaws preventing them from detecting even straightforward instances of crypto-API misuse found in applications. On the other hand, ChatGPT’s accurate response to such test cases raises a curiosity to test whether LLMs can effectively detect cryptographic misuses better than these static analysis tools. To extend it, novice developers intrigued by the power of LLMs capabilities, have started adapting to the technology by using such platforms to look for coding solutions. With the increasing use of AI to look for code solutions by novice developers, little research is done in this area to explore whether prominent platforms, such as ChatGPT, are effective in detecting these cryptographic misuses. It will also likely conclude whether the code provided by these LLMs for cryptographic problems is secure or not. This insight guides our approach to systematically compare widely known crypto detectors with LLMs.

Although early analysis showed promising results for Open AI’s ChatGPT, an intensive study was conducted to conclude the effectiveness of LLMs in detecting cryptographic misuses. LLMs can generate and understand code in various programming languages, however, static tools and benchmarking datasets used for comparison in our study rely only on Java. Therefore, the scope of our work is limited to Java as a programming language.

\section{Background}
Recently, security researchers have expressed significant interest in externally validating static analysis tools~\cite{b12,b13}. Specifically, there's a growing recognition that while static analysis security tools are theoretically sound, they can be ``soundy'' in practice. This means they consist of a core set of sound decisions but also include certain strategically unsound choices made for practical reasons, such as performance or precision considerations~\cite{b23}. In this section, we provide a brief overview of the static analysis tools that will be used for comparison with LLMs. Moreover, we also discuss similar work done in this domain so far.

\textit{CryptoGuard}~\cite{b8,b24} expands on Soot ~\cite{b24}, a widely employed program analysis framework, for the static analysis of Java bytecode~\cite{b25}. It specifically addresses vulnerabilities outlined in CWE-327, CWE-295, CWE-330, CWE-326, CWE-798, and CWE-757. To attain high precision in vulnerability detection, CryptoGuard employs both backward and forward slicing in a context- and field-sensitive manner. Nevertheless, since eight of CryptoGuard's rules pertain to constant value usage, blindly applying existing slicing techniques may inaccurately flag constants covered by program slices but unrelated to security. To address this, CryptoGuard introduces refinement algorithms that eliminate false alarms based on cryptography domain knowledge.

\textit{CogniCrypt}~\cite{b9} assists developers in appropriately utilizing APIs through two primary methods. Firstly, for common tasks like data encryption, CogniCrypt generates code based on high-level task descriptions provided in English. Secondly, CogniCrypt employs rules specified in a domain-specific language (DSL) called CrySL~\cite{b26} to identify API misuses. CogniCrypt translates CrySL rules into context-sensitive, flow-sensitive, and demand-driven static analysis, enabling users to enhance the tool's capabilities by creating new rules. In case of identified API misuse, CogniCrypt provides guidance on its resolution, such as substituting an insecure parameter value with a secure one. Its pattern set pertains to CWE-327, CWE-295, CWE-330, CWE-326, CWE-798, and CWE-757.

\textit{Snyk Code}~\cite{b22}, is a tool developed by Snyk that focuses on static application security testing (SAST). It helps developers identify and fix vulnerabilities in their code. Snyk Code integrates with various development environments and continuously scans codebases for security issues, providing actionable insights and recommendations to remediate detected vulnerabilities. Snyk Code stands out because of its developer-friendly approach, offering integrations with popular IDEs, CI/CD pipelines, and repositories. It uses machine learning and semantic analysis to detect issues, ensuring high accuracy and relevance in its findings. The tool supports multiple programming languages and frameworks, making it versatile for various development needs.

Static analysis tools often suffer from a high rate of false positives, flagging issues that do not pose a threat and creating a disconnect between reported misuse alerts and real vulnerabilities. Chen et al.~\cite{b27} examine these limitations by analyzing the rules, models, and implementations of such tools, highlighting the need for improvements in their precision and usability. In contrast, the use of LLMs for code analysis is gaining traction, as demonstrated by Fang et al.~\cite{b28}, who concluded that advanced LLMs like ChatGPT 3.5 and 4.0 offer promising capabilities for more accurate code review. This suggests that LLMs might mitigate the issue of false positives prevalent in static tools, providing a potentially safer and more effective alternative for detecting vulnerabilities.

\begin{table*}[t]
\centering
\renewcommand{\arraystretch}{1.3} 
\begin{tabular}{@{\extracolsep{\fill}} l c c c c c @{}}
\hline
\textbf{Name} & \multicolumn{3}{c}{\textbf{Benchmarks}} & \textbf{LLM-based Detection} & \textbf{Evaluating Quality of LLM Response} \\ 
\cline{2-4}
& \textbf{CryptoAPI} & \textbf{MASC} & \textbf{OWASP} & & \\ 
\hline
Firouzi et al.~\cite{b33} & \checkmark & \texttimes & \texttimes & \checkmark & \texttimes \\ 
Xia et al.~\cite{b34} & \checkmark & \checkmark & \texttimes & \checkmark & \texttimes \\ 
Our Work & \checkmark & \checkmark & \checkmark & \checkmark & \checkmark \\ 
\hline
\end{tabular}
\caption{Overview of Existing Work on LLM-Based Detection}
\label{Existing_Work_Comparison}
\end{table*}

Liu et al.~\cite{b29} highlight the promising potential of using ChatGPT for vulnerability management. Their research showed that ChatGPT is notably effective in tasks such as generating titles for software bug reports. Similarly, with the rise in popularity of LLMs, researchers have started exploring the domain of detecting cryptographic misuses using LLMs~\cite{b33,b34}. However, no prior work has examined the extent to which the text-based responses of LLMs assist developers in fixing misuse instances.

Despite the progress made by static analysis tools in detecting cryptographic misuse, significant gaps remain in their coverage and adaptability to varied misuse patterns. Existing tools often lack robustness when faced with mutated misuse instances, limiting their effectiveness in real-world scenarios. Our study addressed these gaps by conducting a comprehensive evaluation of static tools against LLMs using the OWASP and CryptoAPI benchmarks. Additionally, we tested the robustness of LLMs with mutated misuse cases, assessing their adaptability beyond typical patterns. To further assist developers, we introduced a keyword-based approach to measure the actionability of LLM responses and specificity to evaluate whether LLMs provide relevant examples to aid in code correction. By addressing these critical aspects, our work explored the potential of replacing static tools with LLMs to assist developers in writing secure code. Table~\ref{Existing_Work_Comparison} provides an overview of existing work in LLM-based detection, highlighting the unique aspects and contributions of our work compared to prior research.

\section{Methodology}
This section describes our methodology for evaluating LLMs in cryptographic misuse detection. We assessed LLMs against various datasets and further analyzed their responses to determine their effectiveness in detecting misuse instances.

\subsection{Selection Criteria}
\subsubsection{Static Tools}
A range of tools exists in both literature and industry for detecting cryptographic vulnerabilities in Java code, as most prior research has primarily focused on this language. We shortlisted two tools from academia and one from industry. For academia, we focused on tools that accept input in the form of a JAR file and are generic rather than specific to a particular framework. As a result, tools specific to Android applications were excluded from this study. We also prioritized tools that have gained popularity in the literature to enhance the credibility of our work when comparing them with LLM models. We selected Cryptoguard~\cite{b8} and CogniCrypt SAST~\cite{b9} as our static tools from academia. Both Cryptoguard and CogniCrypt were set up separately on an Ubuntu VM to run test cases through the terminal. Cryptoguard requires Java 8, while CogniCrypt requires at least Java 11 for setup. For the industry-related tool, we chose Snyk Code~\cite{b22}, a popular option for comparison with other static tools and LLMs. Snyk Code was installed as an extension for Visual Studio Code. 
Table~\ref{Version_Static_Tools} mentions the static analysis tools along with their versions. The results from all the tools were extracted as JSON files, which were later analyzed against the benchmark labels to obtain the results.

\begin{table}[ht]
    \centering
    \begin{tabular}{@{}ll@{}}
        \toprule
        Tool      & Version         \\ \midrule
        CryptoGuard        & 04.05.03                \\
        CogniCrypt         & 3.0.2 and 4.0.1         \\
        Snyk Code          & 2.18.2                  \\ \bottomrule
    \end{tabular}
    \caption{Version of Static Analysis Tools}
    \label{Version_Static_Tools}    
\end{table}

\subsubsection{Benchmark Datasets}
To assess the tools' effectiveness, we conducted thorough online searches for open-source third-party benchmarks that categorize programs according to their accurate or erroneous utilization of cryptographic APIs. We selected three datasets: CryptoAPI Bench~\cite{b20}, OWASP Benchmark~\cite{b21}, and MASC dataset~\cite{b13}, widely used in existing research. These datasets contain test cases to test out the cryptographic misuses for Java programming languages.

\begin{itemize}
    \item CryptoAPI Bench~\cite{b20} consists of a total of 182 test cases, of which 181 are labeled and one unlabeled. Consequently, the unlabeled test case was excluded from the testing. Each labeled test case indicates whether it represents a true misuse case or a false misuse case. Specifically, 144 of these cases are true misuse cases, while 37 are false misuse cases.
    \item The OWASP Benchmark~\cite{b21}, serves as a Java test suite aimed at evaluating the efficacy of automated vulnerability detection methods. This benchmark consolidates vulnerabilities that have been recently identified in the Common Weakness Enumeration (CWE)~\cite{b31} which is endorsed as a suitable evaluation dataset for Application Security Testing tools. We downloaded version 1.2 of this benchmark, which contains a total of 2,740 Java programs. Given that not every program employs security APIs, our analysis was limited to three specific categories: weak cryptography, weak hashing, and weak randomness. As a result, our study includes 975 programs from the original dataset, comprising 477 programs that exhibit labeled misuse of security APIs and 498 programs that illustrate proper usage.
    \item The MASC dataset, designed by Ami et al.~\cite{b13}, applies mutation testing to identify vulnerabilities in crypto detectors, assessing their ability to withstand code modifications. We manually selected 30 test cases from MASC’s minimal test suites, targeting five prevalent flaw types in modern crypto detectors.
\end{itemize}

The outline of test cases for the benchmarking datasets is shown in Table~\ref{Benchmark_Labels}. We chose to employ existing benchmark datasets rather than creating new ones for two main reasons. Firstly, these benchmarks are publicly accessible and have been carefully curated by diverse groups of stakeholders, ensuring that our empirical comparison is both comprehensive and easily replicable by others. Secondly, specific benchmarks like OWASP Benchmark are widely recognized and hold substantial influence in the industry. The use of these datasets on static analysis tools along with LLMs concluded our finding for RQ1. Additionally, testing out LLMs with mutations helped us address RQ2.

\begin{table}[ht]
    \centering
    \begin{tabular}{@{}lccc@{}}
        \toprule
        Benchmarking Dataset & True Labels & False Labels & Total \\ \midrule
        CryptoAPI Bench               & 144                  & 37                    & 181            \\
        MASC                         & 29                  & 1                   & 30 \\
        OWASP                         & 477                  & 498                   & 975		\\\bottomrule
    \end{tabular}
    \caption{Labels of Benchmarking Datasets}
    \label{Benchmark_Labels}
\end{table}

\subsubsection{Large Language Models}
Of the popular LLM models, OpenAI’s GPT and Google’s Gemini were selected for their advanced capabilities in converting text to code and identifying vulnerabilities through sophisticated code analysis features. For each misuse instance, LLMs were provided with a standardized prompt (see Appendix~\ref{sec:LLM_Prompt} for details). This prompt was designed to include details comparable to those captured by other static analysis tools, such as the method name, starting line number, highlighted message, message description (including the reason for misuse), the lines of code containing the misuse, and a label indicating whether the code involved cryptographic misuse. In all cases, the LLM response followed a similar pattern, providing the label indicating whether there was cryptographic misuse, the name of the method containing the misuse, the starting line number, the line of code, a highlighted message for the misuse, and corresponding details explaining the cause of the misuse along with best practices to follow. The LLMs and their respective versions are listed in Table~\ref{LLM_versions}.

\begin{table}[ht]
    \centering
    \begin{tabular}{@{}llll@{}}
        \toprule
        LLM & Version  & Release Date \\ \midrule
        Open AI’s GPT              &    gpt-4o-mini-2024-07-18 & July 18, 2024        		   \\
        Google’s Gemini                         &       gemini-1.5-flash-002 & September 24, 2024     \\ \bottomrule
    \end{tabular}
    \caption{Version of Large Language Models}
    \label{LLM_versions}
\end{table}

\subsection{Evaluation Metrics}
To evaluate the static tools with the LLMs, the accuracy of each static tool and LLMs will be measured by these 7 metrics: \textbf{\textit{True Positive, False Positive, False Negative, True Negative, Precision, Recall, and F-score}}.\\

\subsubsection{LLM Specific Metrics}
 Since Large Language Models generate text, traditional performance metrics such as precision, recall, and F-score are insufficient for fully assessing their capabilities. Therefore, in addition to these metrics, we employed two measures influenced by Redmiles et al.~\cite{b30} to evaluate security and privacy advice: \textit{\textbf{Actionability}} and \textit{\textbf{Specificity}}.\\
To evaluate these metrics, we utilized specific keywords proposed by Rostami et al.~\cite{b35} to deduce actionability in information security policies. We utilized their proposed list of keywords to evaluate whether the response from LLM is actionable. The keywords are listed in Table~\ref{actionable_keywords}. For instance, if the text contains words such as ``should’’, ``need”, or ``must’’, we categorize it as ``Actionable’’. Furthermore, a response is considered higher ``Specificity” if it includes a link to a resource (e.g., a URL) or CWE. 

\begin{table}[ht]
\centering
\begin{tabular}{l}
\toprule
Keywords \\ 
\midrule
never \\
need \\
should \\
not \\
forbidden \\
must \\
shall \\
\bottomrule
\end{tabular}
\caption{Actionable Keywords}
\label{actionable_keywords}
\end{table}

For example, one of the mutated test cases replaces AES with DES by replacing ``A" with ``D". The misuse line is as follows:

\textit{Cipher c = Cipher.getInstance(``AES".replace(``A", ``D''));}\\

For this particular case, GPT gave the misuse description as:\\ 
\textit{``The usage of ``AES".replace(``A", ``D")` is a significant misuse. This expression incorrectly modifies the string that is supposed to define the algorithm name, changing it from "AES" to "DES". Since "DES" is a less secure algorithm, this can lead to vulnerabilities in the encryption implementation realized by this code. This does \textbf{not} represent a proper way to specify or select cryptographic algorithms, as it can easily be manipulated or misunderstood."}

The presence of an actionable keyword ``not" in GPT's response provides clear feedback to developers to fix this misuse instance.

In contrast, the absence of actionable keywords in an LLM response may limit developers’ ability to understand and address the misuse effectively. For instance, GPT provided the following response in another case where AES was replaced by DES:\\
\textit{``The issue arises because the program allows for the selection of weak ciphers (like DES) through method chaining without any safeguards or validation mechanisms to ensure that only strong ciphers are used. This directly impacts the security posture of the application by making it susceptible to various cryptographic attacks, thereby constituting a cryptographic misuse."}

Here, the lack of actionable keywords may limit the guidance that novice developers need, reducing the response’s effectiveness in directing them toward an appropriate fix and failing to emphasize the potential consequences of the misuse.

\begin{figure}[htbp]
    \centering
    \includegraphics[width=0.48\textwidth]{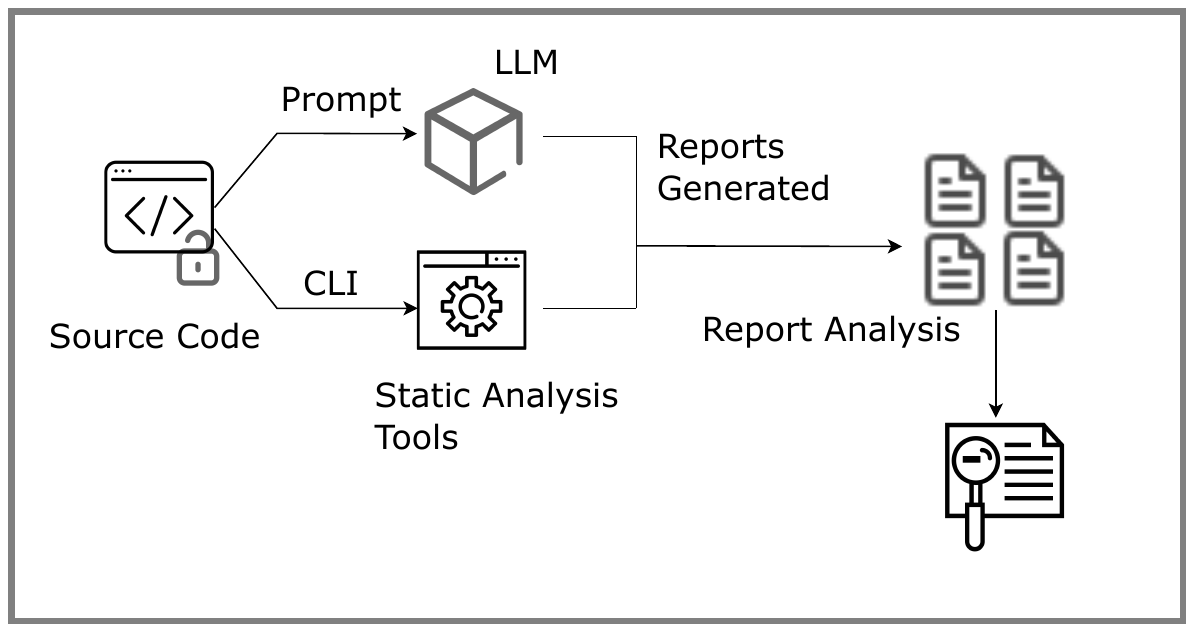}
    \caption{Evaluation Methodology for Comparing Static Tools with LLMs}
    \label{Figure1}
\end{figure}

\subsection{Experiments}
Figure~\ref{Figure1} shows our evaluation methodology for LLM-based cryptographic misuse detection. The methodology is divided into three sections: LLM vs Static Tools, GPT vs Gemini, and Evaluating LLM responses.

\subsubsection{LLM vs Static Tools }
We began our experiments by testing LLMs and static tools with the CryptoAPI and OWASP benchmarks to evaluate their effectiveness in detecting cryptographic misuses, using the evaluation metrics outlined in the metrics section. This helped us understand how well each approach identifies cryptographic misuses and allowed us to compare the strengths of LLMs with those of static tools. This contributed to our findings on RQ1.

\subsubsection{GPT vs Gemini}
In the second phase of our experiments, we tested GPT and Gemini using mutated test cases from the MASC dataset to evaluate their effectiveness in detecting cryptographic misuses. This allowed us to assess the robustness of these LLMs in identifying cryptographic misuse in varied and altered misuse scenarios. This concluded our findings for RQ2.

\subsubsection{Evaluating LLM Responses }
Additionally, we analyzed the responses of LLMs using the Actionability and Specificity metrics. If the response from LLM has a keyword from the Table~\ref{actionable_keywords}, it was considered actionable. The number of actionable responses in a particular dataset gives the total actionability of LLM for that dataset. To determine the Specificity of the LLM response, we manually checked each response to see if the LLM provided a link or referenced a CWE that could assist developers. If the response of LLM has a link or CWE reference, it was considered specific. The number of test cases with specific responses in a particular dataset provides the specificity of LLM and is compared with the other LLM to compare the specificity.

This comprehensive approach offered a clearer understanding of the responses provided by LLMs, assessing whether each response qualified as actionable advice for developers. Additionally, it revealed whether any links provided by the LLMs effectively guide developers to relevant resources for addressing specific cryptographic misuses, thus contributing insights for RQ3.

\section{Results and Findings}
In this section, we present the results of our work on detecting cryptographic misuses for CryptoAPI and OWASP benchmarks, gathered using the selected static tools and GPT. Furthermore, we extended the results by comparing GPT and Gemini on the MASC dataset to determine which LLM performed better. We also analyzed the responses from both LLMs to gain insights into which one offered more actionable and specific advice.

\begin{table*}[htbp]
\centering
\setlength{\tabcolsep}{6pt}
\renewcommand{\arraystretch}{1.5}
\begin{tabular}{llccccccc}
\toprule
Benchmark & Tool & True Positive & False Positive & False Negative & True Negative & Precision (\%) & Recall (\%) & F Score (\%) \\ 
\midrule
\multirow{5}{*}{{CryptoAPI}} 
    & GPT 4-o-mini    & \textbf{141} & 37 & \textbf{3}  & 0  & 79.2 & \textbf{97.9} & \textbf{87.6} \\ 
    & CryptoGuard         & 120 & \textbf{19} & 24 & \textbf{18} & \textbf{86.3} & 83.3 & 84.8 \\ 
    & CogniCrypt 4.0.1    & 113 & 28 & 31 & 9  & 80.1 & 78.5 & 79.3 \\ 
    & CogniCrypt 3.0.2    & 110 & 31 & 34 & 6  & 78.0 & 76.4 & 77.2 \\ 
    & Snyk Code           & 93  & 30 & 51 & 7  & 75.6 & 64.6 & 69.7 \\ 
\midrule
\multirow{5}{*}{{OWASP}} 
    & CryptoGuard         & 437 & \textbf{27} & 40 & \textbf{471} & \textbf{94.2} & 91.6 & \textbf{92.9} \\ 
    & GPT 4-o-mini    & \textbf{477} & 498 & \textbf{0} & 0   & 48.9 & \textbf{100.0} & 65.7 \\ 
    & Snyk Code           & 477 & 498 & \textbf{0} & 0   & 48.9 & \textbf{100.0} & 65.7 \\ 
    & CogniCrypt 4.0.1    & 259 & 200 & 218 & 298 & 56.4 & 54.3 & 55.3 \\ 
    & CogniCrypt 3.0.2    & 259 & 475 & 218 & 23  & 35.3 & 54.3 & 42.8 \\ 
\bottomrule
\end{tabular}
\caption{Evaluation Metrics for CryptoAPI and OWASP Benchmark}
\label{Evaluation_Metrics_for_Benchmarks}
\end{table*}

\subsection{LLMs vs Static Tools}
\subsubsection{CryptoAPI Benchmark}
CryptoAPI benchmark includes a total of 181 test cases, each labeled to indicate whether it represents a true misuse case or not. Specifically, 144 of these cases are true misuse cases, while 37 are false misuse cases. The performance and effectiveness of each tool in identifying cryptographic misuses are detailed below, offering insights into their strengths and limitations in practical applications.

For the CryptoAPI benchmark, GPT 4-o-mini achieved the highest F-score of 87.6\%, with an impressive recall rate of 97.9\%, accurately detecting 141 out of 144 misuse cases. CryptoGuard followed closely with an F-score of 84.8\% and a recall of 83.3\%, missing 24 misuse cases. The detailed performance metrics for each tool are presented in Table~\ref{Evaluation_Metrics_for_Benchmarks}.

\begin{figure}[htbp]
    \centering
    \includegraphics[width=\linewidth]{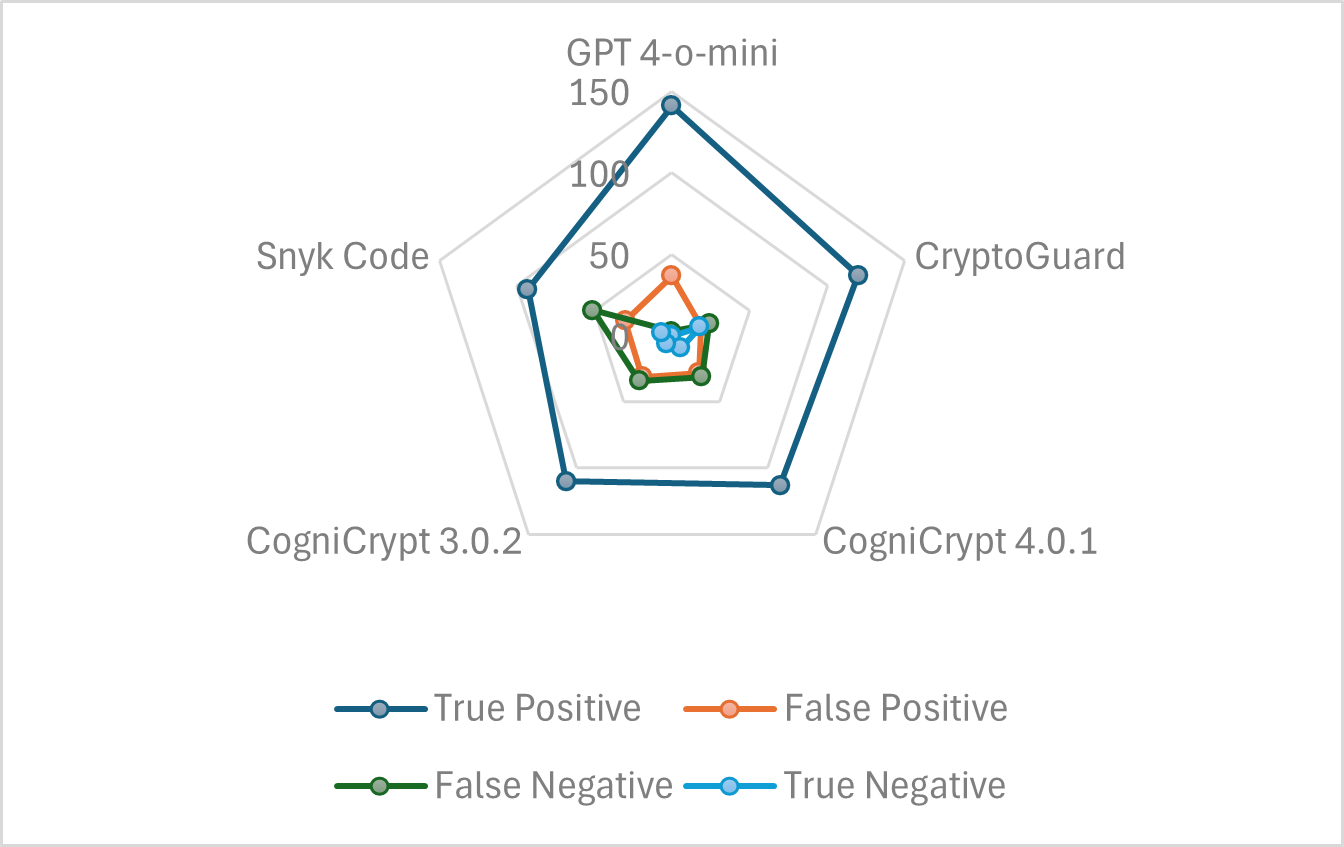} 
    \caption{Tools Comparison on CryptoAPI Benchmark}
    \label{CryptoAPI_Radar}
\end{figure}

\begin{figure}[ht]
    \centering
    \includegraphics[width=\linewidth]{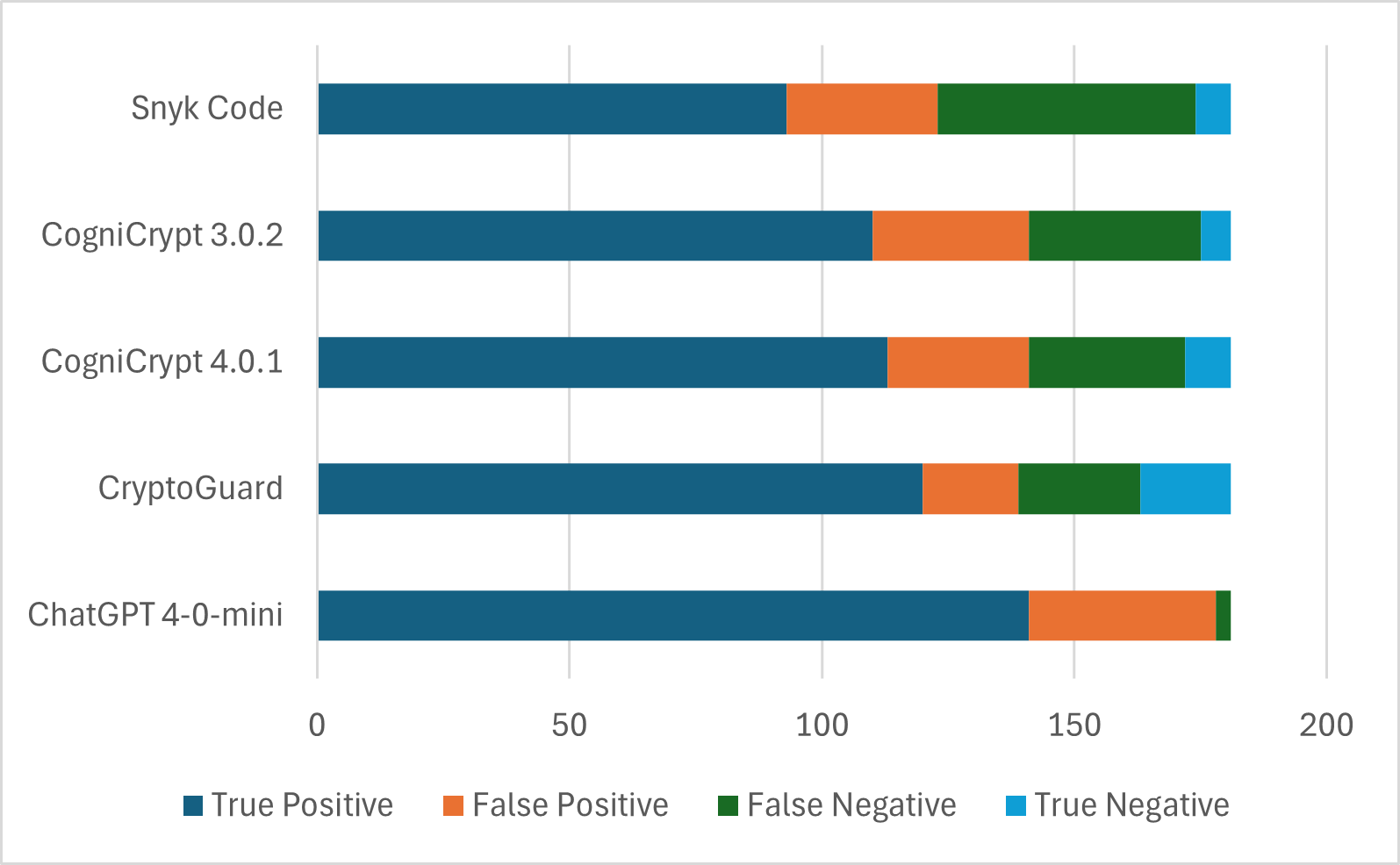} 
    \caption{Detection Results for CryptoAPI Benchmark Across Different Tools}
    \label{CryptoAPI_Bar}
\end{figure}

GPT 4-o-mini’s strong accuracy in detecting misuse cases highlights its potential as an effective alternative to specialized static analysis tools for identifying cryptographic misuses. Unlike traditional tools that rely on fixed rule sets and heuristics—effective mainly for straightforward cases—GPT 4-o-mini uses advanced pattern recognition and contextual adaptability, enabling it to handle complex or ambiguous misuse cases with high precision and recall, though it has a slightly higher rate of false positives. The key metrics of the tools are highlighted in Figure~\ref{CryptoAPI_Radar} and Figure~\ref{CryptoAPI_Bar}. This flexibility allows GPT 4-o-mini to excel where static tools like CryptoGuard and CogniCrypt, limited by their predefined rule coverage and capabilities, may miss specific misuse types, such as Java’s Secret Key Factory or Hostname Verifier API cases.

\begin{tcolorbox}[colback=lightgray, colframe=black, boxrule=0.5pt, arc=0mm, left=1mm, right=1mm, top=1mm, bottom=1mm, fonttitle=\bfseries]
\textbf{Finding 1 RQ1:} GPT 4-o-mini achieved the highest F-score (87.6\%) among tools for CryptoAPI benchmark, showing strong cryptographic misuse detection despite not being specialized, followed closely by CryptoGuard. Snyk Code and CogniCrypt had higher error rates, making them less reliable overall.
\end{tcolorbox}

\subsubsection{OWASP Benchmark}
The OWASP benchmark consists of 975 test cases, each labeled to indicate whether it represents a true misuse case. Of these, 477 are true misuse cases, while 498 are false misuse cases. The performance and effectiveness of each tool in identifying cryptographic misuses are discussed, offering insights into which tool performed best against the benchmark.

CryptoGuard achieved the highest F-score of 92.9\%, with a precision of 94.2\% and a recall of 91.6\%, indicating its ability to accurately detect true misuse cases while keeping false positives low. Snyk Code and GPT 4-o-mini both followed with an F-score of 65.7\%. Although these tools had perfect recall, their high number of false positives reduced their overall accuracy, revealing strong detection capabilities but limited precision. The OWASP benchmark results are summarized in Table \ref{Evaluation_Metrics_for_Benchmarks}.

\begin{figure}[ht]
    \centering
    \includegraphics[width=\linewidth]{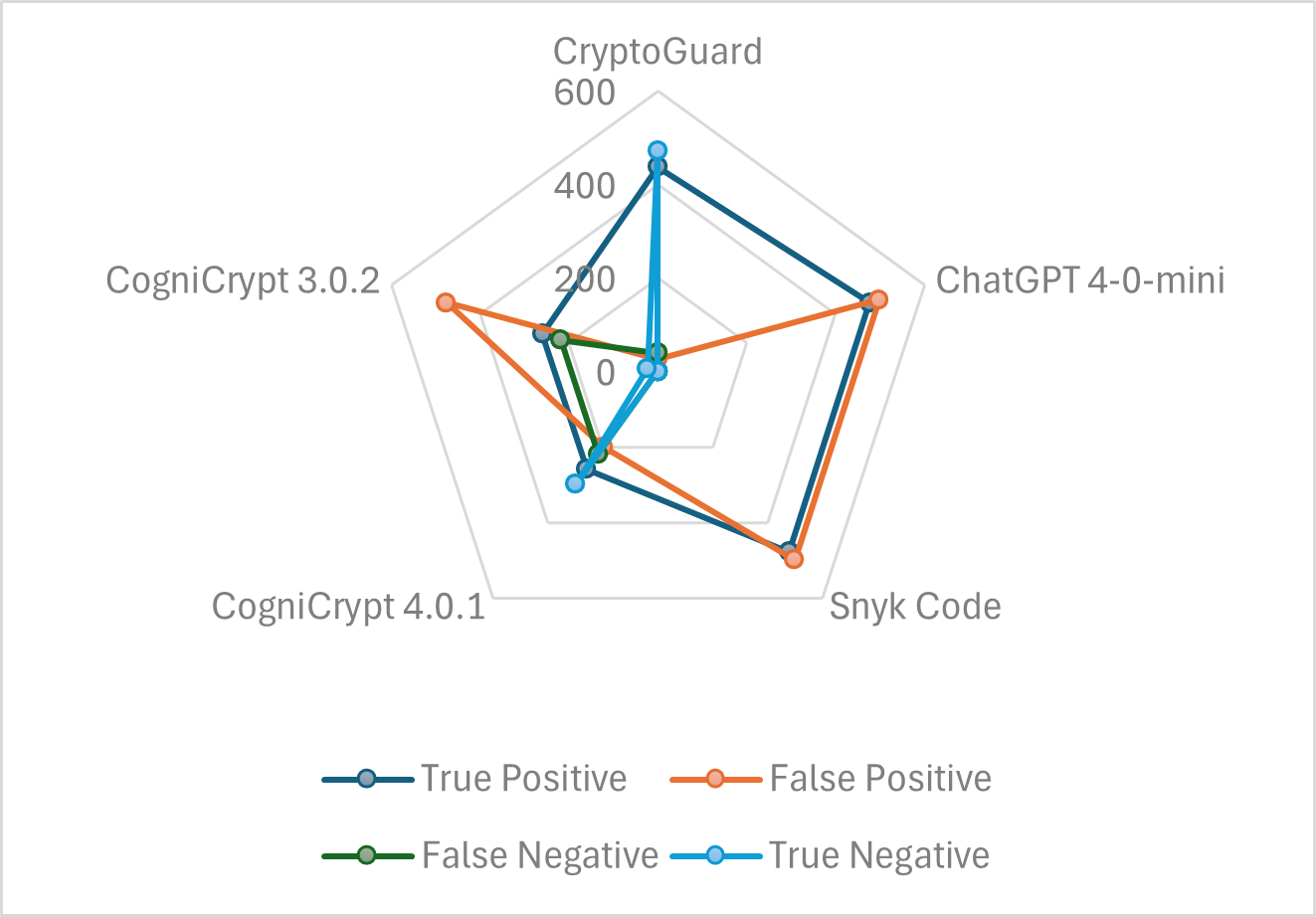} 
    \caption{Tools Comparison on OWASP Benchmark}
    \label{OWASP_Radar}
\end{figure}

\begin{figure}[ht]
    \centering
    \includegraphics[width=\linewidth]{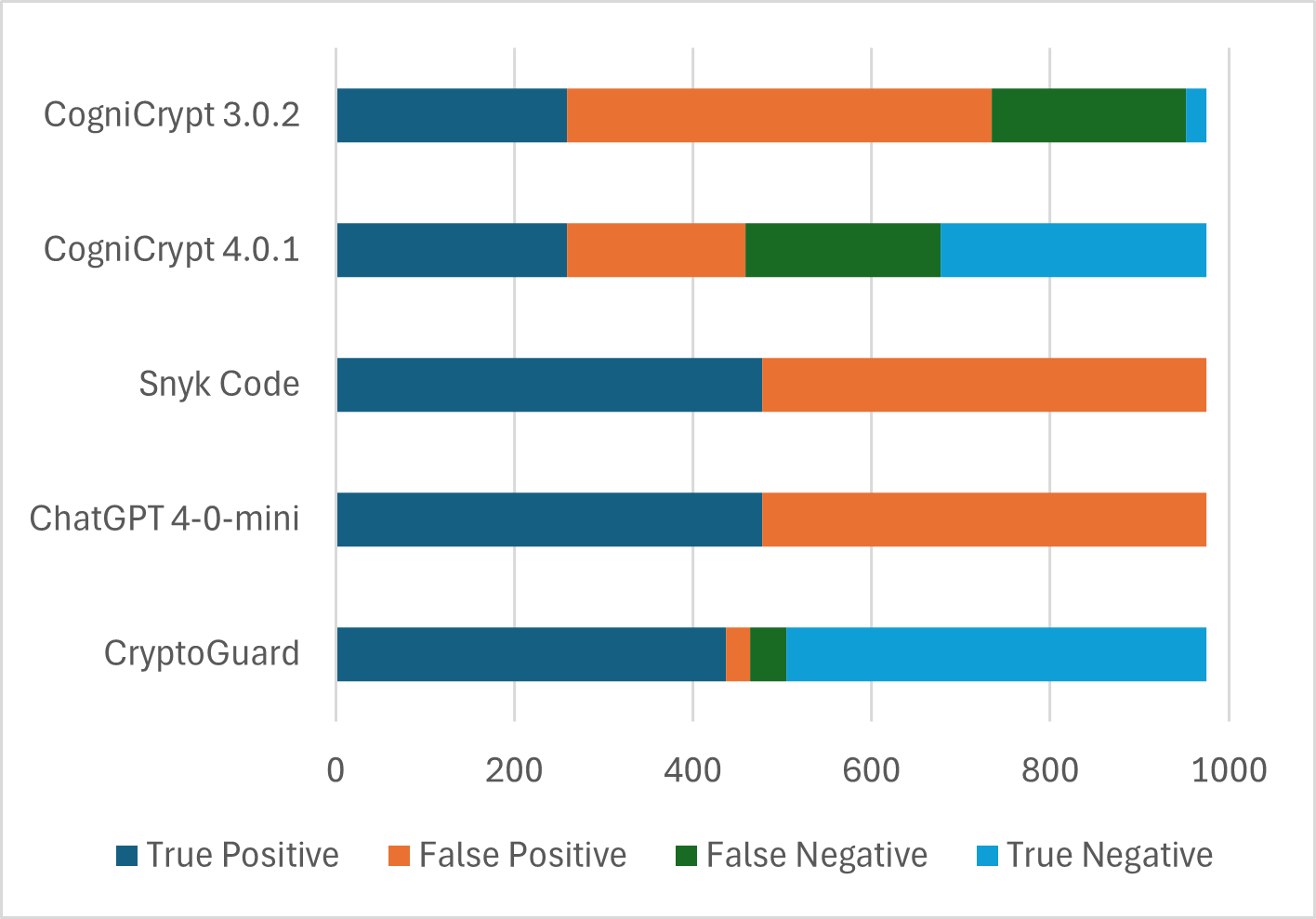} 
    \caption{Detection Results for OWASP Benchmark Across Different Tools}
    \label{OWASP_Bar}
\end{figure}

GPT 4-o-mini’s high false positive rate can be attributed to its text-based responses, which often suggest best practices even when no misuse is present. This tendency led GPT 4-o-mini to label cases as ``TRUE” for best practice recommendations, even without actual cryptographic misuse, lowering its precision in distinguishing true misuse cases. Additionally, the OWASP test cases had a higher average line count than the CryptoAPI test cases, resulting in more code for GPT to analyze. With longer code samples, GPT tended to provide general recommendations rather than focusing solely on detecting cryptographic misuses, often labeling false cases as true. Conversely, GPT performed more accurately with smaller code samples compared to test cases with more lines of code. Figure~\ref{OWASP_Radar} and Figure~\ref{OWASP_Bar} display the key metrics of the tools tested on the OWASP Benchmark, highlighting CryptoGuard’s superior performance over other tools.

\begin{tcolorbox}[colback=lightgray, colframe=black, boxrule=0.5pt, arc=0mm, left=1mm, right=1mm, top=1mm, bottom=1mm, fonttitle=\bfseries]
\textbf{Finding 2 RQ1:} CryptoGuard outperforms other tools on the OWASP benchmark with a 92.9\% F-score, while GPT and Snyk Code lag behind at 65.7\%, demonstrating CryptoGuard's higher accuracy in detecting cryptographic misuses.
\end{tcolorbox}

\subsubsection{Time Comparison}

In addition to comparing the results of each static tool with GPT, we also compared the time it took to run both benchmarks. The tools were configured on an Ubuntu virtual machine (VM) hosted on a laptop. The laptop's specifications are: 1) Operating System: Windows 11, 2) CPU: i7-1255U at 1.70 GHz, 3) RAM: 16 GB. The VMs to set up tools were created with the following specifications: 1) Operating System: Ubuntu, 2) RAM: 10 GB, 3) JVM heap size: 8 GB, and 4) Number of Processors: 4.

\begin{table}[ht]
\centering
\setlength{\tabcolsep}{10pt}
\renewcommand{\arraystretch}{1.5}
\begin{tabular}{lll} 
\toprule
Tools & CryptoAPI & OWASP \\ 
\midrule
CryptoGuard       & 1m 48s & 10m 59s \\ 
CogniCrypt 3.0.2  & 54s    & 16m 35s \\ 
CogniCrypt 4.0.1  & 1m 7s  & 23m 26s \\ 
Snyk Code         & \textbf{8s}     & \textbf{52s} \\ 
GPT 4-o-mini      & 8m 45s & 71m \\ 
\bottomrule
\end{tabular}
\caption{Execution Time of Tools for CryptoAPI and OWASP Benchmarks}
\label{Time_Taken_by_Tools}
\end{table}

Test cases from datasets were run on GPT using an API and the time it took to get responses for each benchmark was recorded. The static analysis tools, including CryptoGuard and CogniCrypt, were run using the CLI command, whereas Snyk code was utilized as an extension to Visual Studio Code. The results of time taken for each tool for both benchmarks are highlighted in Table~\ref{Time_Taken_by_Tools}. 

The data indicates that Snyk Code is the fastest tool across both the CryptoAPI and OWASP benchmarks, with execution times of 8 seconds and 52 seconds, respectively. CryptoGuard also performs efficiently, especially on the CryptoAPI benchmark (1m 48s), though it takes longer on the OWASP benchmark (10m 59s). In contrast, GPT 4-o-mini has the longest execution times, requiring 8m 45s for the CryptoAPI and 71 minutes for the OWASP benchmark, highlighting a notable delay compared to other tools. This suggests that while GPT 4-o-mini may excel in detection accuracy, its slower processing speed could limit its scalability for large datasets.

\begin{tcolorbox}[colback=lightgray, colframe=black, boxrule=0.5pt, arc=0mm, left=1mm, right=1mm, top=1mm, bottom=1mm, fonttitle=\bfseries]
\textbf{Finding 3 RQ1:} Snyk Code exhibits the fastest execution time across both benchmarks, followed by CryptoGuard with the second-best performance. In contrast, GPT 4-o-mini, despite its detection strengths, has significantly slower processing speeds, which may hinder its scalability for larger code bases.
\end{tcolorbox}

\subsection{Comparing LLMs with Datasets}
For this section, we evaluated the performance of two LLMs, GPT 4-o-mini and Gemini 1.5 Flash, against the benchmark datasets. We also presented the results of running mutated test cases to determine which model performed better. GPT 4-o-mini was accessed via an API, while Gemini 1.5 Flash was accessed through its chat interface. For the CryptoAPI and OWASP benchmarks, Gemini was tested on 30 random test cases, and the scores were compared with GPT to determine which performed better. Similarly, both LLMs were tested on 30 manually curated MASC test cases for GPT and Gemini to conclude which performed better. Examples of some LLM responses are provided in Appendix~\ref{sec:LLM_Responses}.

To compare GPT and Gemini, 30 test cases from the CryptoAPI and OWASP benchmarks were randomly selected, with Gemini evaluated using the same prompts as GPT. On the CryptoAPI benchmark, GPT achieved a recall of 97.9\% and an F-score of 87.6\%, outperforming Gemini’s recall of 87\% and F-score of 83.3\%, though Gemini achieved a slightly higher precision of 80\% compared to GPT’s 79.2\%. For the OWASP benchmark, GPT had an F-score of 65.7\%, while Gemini achieved a perfect recall of 100\% but a lower precision of 46.2\%, resulting in an F-score of 63.2\%. These results suggest that while both models perform similarly on recall, GPT shows greater precision and overall effectiveness in cryptographic misuse detection (see Table~\ref{LLM_comparison_benchmarks}).

\begin{table*}[ht]
\centering
\resizebox{\textwidth}{!}{ 
\begin{tabular}{llccccccc}
\toprule
Benchmark & LLM & True Positive & False Positive & False Negative & True Negative & Precision (\%) & Recall (\%) & F Score (\%) \\

\midrule
\multirow{2}{*}{CryptoAPI} & GPT 4-o-mini         & \textbf{141} & 37  & \textbf{3 }& 0 & 79.2 & \textbf{98}  & \textbf{87.6} \\
                           & Gemini Flash 1.5    & 20  & \textbf{5}   & \textbf{3} & \textbf{2} & \textbf{80.0} & 87  & 83.3 \\
\midrule
\multirow{2}{*}{MASC}      & GPT 4-o-mini         & \textbf{29 } & 1   & \textbf{0} & 0 & 96.7 & \textbf{100} & \textbf{98.3} \\
                           & Gemini Flash 1.5     & 23  & \textbf{0}   & 6 & \textbf{1} & \textbf{100}  & 79.3 & 88.5 \\
\midrule
\multirow{2}{*}{OWASP}     & GPT 4-o-mini         & \textbf{477} & 498 & \textbf{0} & 0 & \textbf{48.9} & \textbf{100} & \textbf{65.7} \\
                           & Gemini Flash 1.5\footnotemark   & \textbf{12}  & \textbf{14 } & \textbf{0} & \textbf{4} & 46.2 & \textbf{100} & 63.2 \\

\bottomrule
\end{tabular}
}
\caption{Comparison of LLMs on Different Benchmarks}
\label{LLM_comparison_benchmarks}
\end{table*}

\footnotetext{Gemini was evaluated against 30 randomly test cases for CryptoAPI and OWASP dataset..}

On the mutation dataset (MASC), which tests robustness, GPT outperformed Gemini by correctly identifying all 29 true misuse cases and flagging the single false positive, resulting in a precision of 96.7\%, a recall of 100\%, and an F-score of 98.3\%. In comparison, Gemini detected 23 of 29 true misuse cases with no additional false positives, yielding a recall of 79.3\% and an F-score of 88.5\%. This demonstrates the strength of LLMs in varied scenarios, an area where many existing static tools still face challenges.

\begin{tcolorbox}[colback=lightgray, colframe=black, boxrule=0.5pt, arc=0mm, left=1mm, right=1mm, top=1mm, bottom=1mm, fonttitle=\bfseries]
\textbf{Finding RQ2:} GPT demonstrates superior performance in detecting cryptographic misuses across all benchmarks, achieving higher recall and F-scores than Gemini. However, this accuracy comes at the cost of a higher rate of false positives. As identifying true misuse is more critical than minimizing false positives, GPT is the more effective LLM for detecting cryptographic misuse.
\end{tcolorbox}

\subsection{Evaluating LLM Responses}
To evaluate the responses of both LLMs, we used two metrics: Actionability and Specificity. Actionability was assessed by examining each LLM response to check for the presence of an actionable keyword. If an actionable keyword was included, the response was marked as actionable; otherwise, it was marked as non-actionable. The percentage of actionable cases for each benchmark was then calculated and compared across LLMs. Similarly, a response was considered specific if it included a reference to the CWE or a link to an external resource; otherwise, it was marked as non-specific. The percentage of specific cases provided a measure of specificity for each benchmark. The results for actionability and specificity across benchmarks are presented in Table~\ref{LLM_Metrics}.

\begin{table*}[htbp]
\small 
\resizebox{1\textwidth}{!}{ 
\centering
\begin{tabular}{llcccccc}
\toprule
Benchmark & LLM & Actionable Responses & Specific Responses & Total Responses & Actionability (\%) & Specificity (\%) \\
\midrule
\multirow{2}{*}{CryptoAPI} & GPT 4-o-mini     & 162  & 0 & 264  & \textbf{61.4} & 0.0 \\
                           & Gemini Flash 1.5 & 9    & 2 & 30   & 30.0 & \textbf{6.7} \\
\midrule
\multirow{2}{*}{MASC}      & GPT 4-o-mini     & 19   & 0 & 30   & \textbf{63.3} & 0.0 \\
                           & Gemini Flash 1.5 & 7    & 7 & 30   & 23.3 & \textbf{23.3} \\
\midrule
\multirow{2}{*}{OWASP}     & GPT 4-o-mini     & 1584 & 0 & 1909 & \textbf{83.0} & 0.0 \\
                           & Gemini Flash 1.5 & 21   & 6 & 30   & 70.0 & \textbf{20.0} \\
\bottomrule
\end{tabular}
}
\caption{Actionability and Specificity of LLMs Across Benchmark Datasets}
\label{LLM_Metrics}
\end{table*}

\begin{figure}[htbp] 
    \centering
    \includegraphics[width=0.48\textwidth]{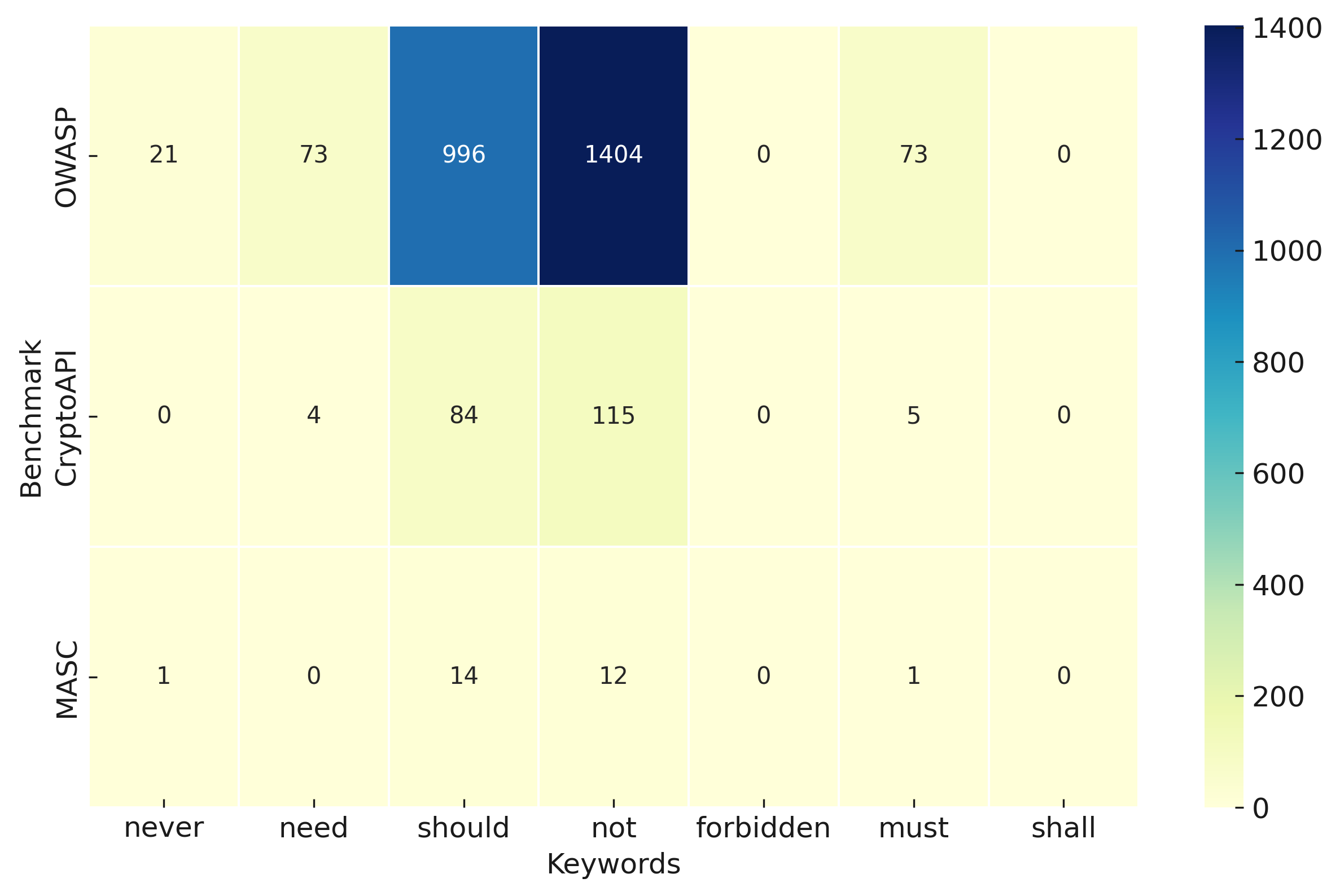}
    \caption{GPT Actionable Keywords Occurence}
    \label{GPT_heatmap}
\end{figure}

\begin{figure}[htbp] 
    \centering
   \includegraphics[width=0.48\textwidth]{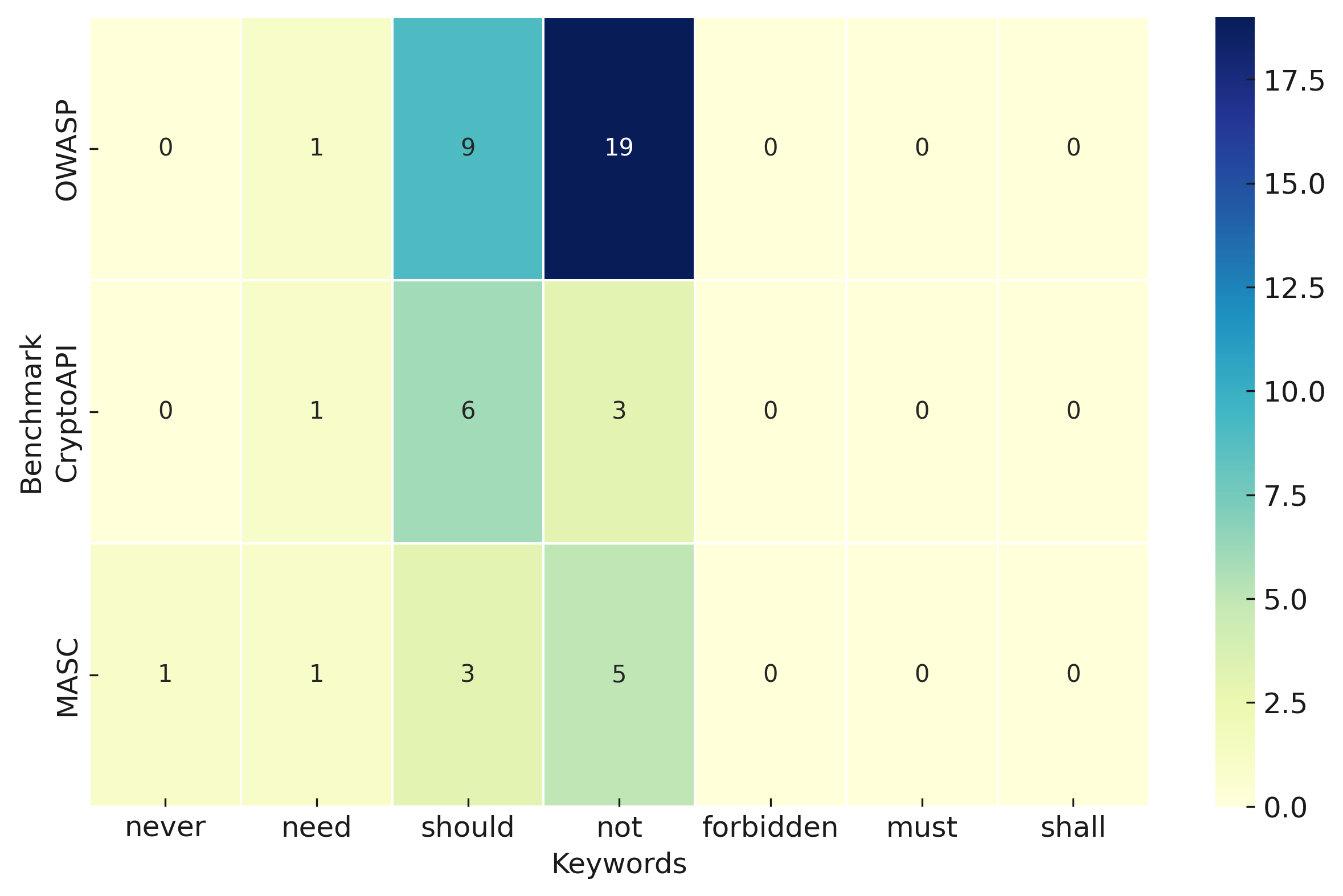}
    \caption{Gemini Actionable Keywords Occurence}
    \label{Gemini_heatmap}
\end{figure}

GPT 4-o-mini produced more actionable responses, with 83.0\% actionability on OWASP and 61.4\% on CryptoAPI, but lacked specificity, scoring 0\% across benchmarks. In contrast, Gemini Flash 1.5 had lower actionability (70.0\% on OWASP and 30.0\% on CryptoAPI) but provided modest specificity, with 20.0\% on OWASP and 23.3\% on MASC. This suggests that GPT 4-o-mini excels in actionability, while Gemini Flash 1.5, though less actionable, offers some specific guidance by referencing CWEs or external resources to fix cryptographic misuses.

Figures~\ref{GPT_heatmap} and~\ref{Gemini_heatmap} show the occurrence of actionable keywords for GPT and Gemini. The most frequently used keywords across benchmarks are ``not" and ``should". In both the OWASP and CryptoAPI benchmarks, ``not" appears frequently, especially in OWASP, with 1404 instances in GPT. ``Should" is also common, with 996 instances in GPT for OWASP, and lower but consistent use across other benchmarks. These keywords highlight a focus on negation and suggestion, guiding actions to avoid or consider. Less frequent keywords, like ``need", ``must", and ``shall", suggest a preference for softer language rather than strict requirements.

\begin{tcolorbox}[colback=lightgray, colframe=black, boxrule=0.5pt, arc=0mm, left=1mm, right=1mm, top=1mm, bottom=1mm, fonttitle=\bfseries]
\textbf{Finding RQ3:} GPT 4-o-mini exhibits high actionability across benchmarks but lacks specificity, while Gemini Flash 1.5, though less actionable, offers some specificity through references to CWEs and external resources.
\end{tcolorbox}

\section{Discussion}
\textbf{F-score vs Recall}
The F-score used to analyze the most effective tool may not be the best measure for evaluating its effectiveness in detecting cryptographic misuses. A key issue with static tools is their tendency to produce a high number of false positives, which significantly impacts the F-score and helps determine which tool performed better. However, for security purposes, recall is the most important measure alongside the F-score, ensuring that the tool does not miss any true misuse instances. Although Snyk Code and GPT have significantly lower F-scores on the OWASP benchmark compared to Cryptoguard, it achieves a perfect recall of 100\%, indicating that it did not miss any true instances. GPT and Snyk Code's conservative approach resulted in a high number of false positives, which impacted its F-score.

While the F1-score is an effective measure for evaluating a tool's performance, it should not be the only criterion for tool selection. In a security context, placing more emphasis on recall will provide more desirable results than relying solely on the F-score. For the OWASP benchmark, calculating the F2-score significantly improves the score of Snyk code and GPT from approximately 66\% to 83\%, highlighting the tool’s effectiveness in detecting true alerts.Table~\ref{F2Score} presents the results for the F2-score, which assigns more weight to recall than precision. 

\begin{table}[ht]
\centering
\small 
\resizebox{0.35\textwidth}{!}{ 
\begin{tabular}{lcc}
\toprule
Tool & F1 Score & F2 Score \\
\midrule
CryptoGuard       & 92.9 & 92.1 \\
GPT 4-o-mini  & 65.7 & 82.7 \\
Snyk Code         & 65.7 & 82.7 \\
CogniCrypt 4.0.1  & 55.3 & 54.7 \\
CogniCrypt 3.0.2  & 42.8 & 49.0 \\
\bottomrule
\end{tabular}
}
\caption{Comparison of F-1 and F-2 Scores for OWASP}
\label{F2Score}
\end{table}

\textbf{Strengths of Large Language Models}
LLMs have an inherent advantage over static analysis tools as static tools rely on a rule-based approach. Any deviations of misuse instance from a rule will likely cause the tool to miss it. However, LLM detection capability is not particular to a specific rule set, and it can detect misuse instances in a variety of code as highlighted by the results of mutated test cases. Similarly, LLMs are not specific to a certain programming language. It can read and understand various programming languages as opposed to static analysis tools that are specific to a programming language. LLMs do not have a framework dependency and now integrations are being supported to adopt LLMs for various frameworks which will likely increase its usage in software development. Even though LLMs currently have a high false positive rate for detecting cryptographic misuses, this is likely to improve as they are trained on more data. Better accuracy than static analysis tools in detecting true alerts for security context shows LLMs can be utilized in this domain.\\

\textbf{Benchmark Inconsistencies}
Our analysis revealed inconsistencies in the CryptoAPI benchmark, where the label sheet, which indicates whether instances are true or false, includes entries for test cases not present in the dataset. Specifically, we found labels for CredentialInStringBBCase2.java and PredictableSeedsABPMCase2.java in the label sheet, although these Java files are missing from the dataset. This mismatch can increase false negatives, potentially giving a misleading impression of the tools' effectiveness. Additionally, the dataset includes Java test cases without corresponding labels, causing tools to detect misuse instances without a reference. To maintain consistency in this study, we disregarded these unlisted test cases. This highlights the need for a comprehensive, standardized dataset to validate tool effectiveness across all Java API categories.

\section{Ecological Validity}
\subsection{Applicability to Real World Scenarios}
Our experiments utilized the OWASP Benchmark, a well-known open-source Java suite that serves as a standard for assessing application security. The CryptoAPI benchmark was also included based on a review of previous studies, where it has been extensively used. Additionally, the MASC dataset was selected to capture the diverse ways in which code can be written by developers. The results obtained from these benchmarks provide insights into the effectiveness of different tools and LLMs in detecting vulnerabilities across various domains. Our focus was specifically on cryptographic misuses, making the results highly relevant for comparing our tool with both industry-standard and academic tools, many of which are evaluated against these benchmarks. Since these benchmarks are commonly used, their results are a reliable measure of a tool’s accuracy in identifying cryptographic misuses.

The increasing adoption of LLMs by novice developers, often in place of traditional search engines, reflects a shift in how coding solutions are sourced. Many coding environments, like Visual Studio Code, now integrate LLM-based assistants such as ChatGPT, allowing developers to prompt these models directly. With LLMs providing quick, context-specific answers, novice developers are more likely to rely on these tools not only for general coding guidance but also for secure coding practices, including cryptographic implementations. This shift means that as more developers turn to LLMs for solutions, they will inherently depend on the security accuracy of these models to prevent cryptographic misuses in their code. Therefore, it becomes critical to evaluate how well LLMs address security contexts, ensuring that the solutions provided are both effective and safe. Understanding this trend highlights the real-world importance of our work, which assesses the quality and security of LLM responses in situations where developers may not be aware of potential misuses. 

\subsection{Threats to Validity}
\textit{Threats to Internal Validity}
The scope of our findings may be limited by the specific tools and datasets chosen for testing. To address this, we plan to expand our evaluation by including additional tools and testing them on examples drawn from real-world cryptographic code. Another potential limitation is data leakage, as the open-source benchmarks used for testing might have been utilized in the training of LLMs, potentially affecting our results; however, the high number of false positives suggests otherwise. In the future, we plan to evaluate test cases by anonymizing class names to prevent any possible data leakage. Additionally, the actionability and specificity metrics evaluated in our study may not fully translate into practical guidance for developers to fix code in every instance.

\textit{Threats to External Validity}
Our results may not generalize beyond the specific tools and datasets used in this study. To address this, we plan to expand our evaluation by incorporating additional tools and real-world cryptographic benchmarks. Additionally, our findings are specific to Java and may not apply in the same way to other programming languages. Moreover, we tested for mutations in test cases that were specific to Java as a programming language, the accuracy of detecting mutant misuses might not apply to other programming languages.

\section{Future Work}
Future research could explore testing LLMs on real-world codebase benchmarks, such as the Apache CryptoAPI benchmark, to assess performance on complex cryptographic scenarios. Expanding the study to include more LLMs, particularly those specialized in code generation, may provide insights into the benefits of using code-focused models over general-purpose ones. Additionally, testing LLMs on benchmarks in other programming languages could reveal language-specific strengths and limitations.

\section{Conclusion}
Our work compared static analysis tools with LLMs, highlighting the strengths of LLMs in detecting cryptographic misuses. However, high false positive rates of LLMs may impact early adoption. Our study focused on the CryptoAPI, OWASP, and MASC datasets, where LLMs demonstrated high accuracy across all benchmarks, often surpassing static tools. We assessed the quality of LLM responses using actionability and specificity metrics. The results of our work suggest that LLMs hold promising potential for use in security analysis.

\section*{Acknowledgement}
We thank the Vulnerability Research Centre at the Communications Security Establishment, Canada, for supporting this research.

\bibliographystyle{IEEEtran} 
\bibliography{references} %

\newcommand{\appendixCaptions}{
    \captionsetup{
        justification=justified,  
        singlelinecheck=false,       
        width=0.5\textwidth,         
        font=footnotesize,           
        skip=5pt ,                    
        format=plain                
    }
}

\appendices
\appendixCaptions

\definecolor{commentgreen}{rgb}{0, 0.5, 0}
\definecolor{keywordblue}{rgb}{0, 0, 0.7}
\definecolor{stringpurple}{rgb}{0.58, 0, 0.82}
\definecolor{lightgray}{rgb}{0.95, 0.95, 0.95}

\lstdefinestyle{javaAppendixStyle}{
  language=Java,
  basicstyle=\ttfamily\small,
  keywordstyle=\color{keywordblue},
  commentstyle=\color{commentgreen},
  stringstyle=\color{stringpurple},
  frame=single,
  numbers=left,
  numberstyle=\tiny,
   xleftmargin=2em,             
  framexleftmargin=2em,        
  tabsize=4,
  breaklines=true,
  captionpos=b
}

\lstdefinestyle{textAppendixStyle}{
  basicstyle=\ttfamily,          
  frame=single,                  
  numbers=left,                  
  xleftmargin=2em,               
  framexleftmargin=2em,          
  breaklines=true,               
  captionpos=b,
  numberstyle=\tiny,
  xleftmargin=2em,             
  framexleftmargin=2em,        
  tabsize=4                      
}

\section{LLM Prompt}
\label{sec:LLM_Prompt}
\setcounter{table}{0}
\renewcommand{\thetable}{\thesection\arabic{table}} 
\renewcommand{\thelstlisting}{\thesection\arabic{lstlisting}} 

Both Large Language Models (LLMs), GPT 4-o-mini and Gemini, received standardized prompts for each test case from the three benchmarks to ensure their responses were consistent. The specific prompt used for each test case is shown in Listing~\ref{lst:LLM_Prompt}. During the experiments, we kept the default settings for model hyper-parameters like temperature, Top P, and frequency penalty to maintain each model’s natural response style. Responses from GPT 4-o-mini were collected automatically through an API, while responses from Gemini were gathered manually from its chat interface.

\begin{lstlisting}[style=textAppendixStyle,caption=Prompt Used to Query LLMs,label=lst:LLM_Prompt]
I want you to detect "Cryptographic misuses" in the given Java code by considering the cryptographic misuse definitions below.

Cryptographic misuses are deviations from best practices while incorporating cryptographic algorithms into your software that could potentially be exploited by an adversary. Broadly, cryptographic misuses can be referred to as bad programming practices that create vulnerabilities and are associated with design flaws and unsafe architectural choices.

Identify the following for each cryptographic misuse and include it in your response. Follow the same template for reporting multiple misuses.

1- Include a label as YES/NO if there are specifically cryptographic misuses in the given code or not.
2- Include the name of the method containing the misuse.
3- Specify a message highlighting the type of misuses. Provide a description in the explanation as to why it is highlighted as a misuse.
4- State the line from the code containing the misuses. Specifically mention the starting line number that is causing misuse.

Code:
// Code for Each test case is placed here

\end{lstlisting}

\section{LLM Responses}
\label{sec:LLM_Responses}
The responses from the LLMs were fairly similar in format due to the structured prompt, though some variations were observed. For instance, in certain cases, the explanation for misuse from GPT 4-o-mini appeared under tags such as ``Misuse Description", while in others, it was labeled as ``Explanation" or a similar term. An example of this can be seen in Listing~\protect\ref{lst:GPT1}, which shows the response for the CryptoAPI Bench test case listed in Listing~\protect\ref{lst:cryptoapi1}. The misuse in this instance involves using the DES algorithm and the ``DES" string to initialize the key.

\begin{lstlisting}[style=javaAppendixStyle, %
caption={CryptoAPI Bench Test Case: BrokenCryptoABMCCase1.java}, %
label=lst:cryptoapi1]
package org.cryptoapi.bench.brokencrypto;
 
import java.security.InvalidKeyException;
import java.security.NoSuchAlgorithmException;
import javax.crypto.NoSuchPaddingException;

public class BrokenCryptoABMCCase1 
{
    public static void main(String[] args) throws NoSuchPaddingException, NoSuchAlgorithmException, InvalidKeyException 
    {
        BrokenCryptoABMC1 bc = new BrokenCryptoABMC1();
        String crypto = "DES/ECB/PKCS5Padding";
        String cryptokey = "DES";
        bc.go(crypto, cryptokey);
    }
}
\end{lstlisting}

\begin{lstlisting}[style=textAppendixStyle,caption={GPT 4-o-mini Response for BrokenCryptoABMCCase1.java},label=lst:GPT1]
Response:
1- **Label**: YES  
2- **Method containing the misuse**: main  
3- **Message highlighting the type of misuses**: Use of weak cryptographic algorithm and mode of operation  
   **Explanation**: The code uses DES (Data Encryption Standard) in ECB (Electronic Codebook) mode, which is considered weak and insecure. DES itself has been deprecated due to its small key size and susceptibility to brute-force attacks, while ECB mode exposes patterns in plaintext since identical plaintext blocks produce identical ciphertext blocks. This can lead to serious vulnerabilities, allowing attackers to perform cryptanalysis or deduce information about the plaintext. Best practices recommend using stronger algorithms (e.g., AES) and secure modes of operation (e.g., GCM or CBC with proper IV management).  
4- **Line from the code containing the misuses**: 10-11 (Lines mentioning `String crypto = "DES/ECB/PKCS5Padding";` and `String cryptokey = "DES";`)
\end{lstlisting}

The results generated by the LLMs across all three benchmarks were compiled and compared with the dataset labels to draw our conclusions. A selection of LLM responses for various misuse instances is presented in Table~\ref{Table_Appendix}. The detailed findings are discussed in Section 5 of this paper.

\clearpage
\begin{sidewaystable}[htbp]
\vspace{8cm}
\centering
\caption{LLM Responses to Test Cases from Different Datasets}
\label{Table_Appendix}
\footnotesize
\begin{tabularx}{\textheight}{ >{\raggedright\arraybackslash}p{1cm} >
{\raggedright\arraybackslash}p{2.8cm} >
{\raggedright\arraybackslash}p{5.5cm} >
{\raggedright\arraybackslash}p{0.7cm} >
{\raggedright\arraybackslash}p{0.5cm} >
{\raggedright\arraybackslash}p{1cm} >
{\RaggedRight\arraybackslash}p{3cm} >
{\RaggedRight\arraybackslash}p{5.4cm}}
\toprule
\textbf{Dataset} & \textbf{Java File} & \textbf{Misuse Instance} & \textbf{Dataset Label} & \textbf{LLM} & \textbf{LLM Label} & \textbf{Message Highlighted} & \textbf{Message Description} \\ 
\midrule
CryptoAPI Bench & BrokenCryptoABICase13 & String crypto = ``RC4"; & TRUE & GPT & YES & Use of insecure cryptographic algorithm & The code uses ``RC4" as the cryptographic algorithm. RC4 is considered insecure due to vulnerabilities that allow for various attacks, including key recovery and plaintext recovery. Best practices recommend avoiding its use in favor of more secure algorithms like AES. By allowing the specification of cryptographic algorithms at runtime, this code opens the door for an adversary to potentially use insecure algorithms deliberately. \\ 
\midrule
CryptoAPI Bench & StaticSaltsABSCase1 & byte[] salt = \{ -94 \}; & TRUE & Gemini & YES & Hardcoded Salt & The code contains a hardcoded salt value, which is a common cryptographic misuse. Hardcoded salts are problematic because they can weaken the security of the cryptographic system. If an attacker discovers the hardcoded salt, they can potentially compromise the security of the system. To address this issue, the salt value should be generated randomly and securely. This can be done using a cryptographically secure random number generator (CSPRNG). Additionally, the salt should be stored securely and not exposed to unauthorized access. \\ 
\midrule
CryptoAPI Bench & BrokenCryptoBBCase5 & Cipher cipher = Cipher.getInstance(``IDEA"); & TRUE & GPT & YES & Use of weak cryptographic algorithm (IDEA) and key size analysis & The method utilizes the IDEA (International Data Encryption Algorithm), which is considered outdated and vulnerable due to its small block size and usage of relatively small key lengths (up to 128 bits). Modern best practices recommend using stronger algorithms such as AES (Advanced Encryption Standard), which offers better security and is more widely supported. \\
\midrule
CryptoAPI Bench & BrokenHashABPSCase3 & md = MessageDigest.getInstance(``SHA-256"); & FALSE & Gemini & NO & N/A & This code snippet does not contain any specific cryptographic misuses based on the provided definition. \\
\midrule
MASC & AES String Replace & Cipher c = Cipher.getInstance(``AES".replace(``A", ``D")); & TRUE & GPT & YES & Misuse of unapproved algorithm specification for Cipher & The usage of \texttt{"AES".replace("A", "D")} is a significant misuse. This expression incorrectly modifies the string that is supposed to define the algorithm name, changing it from ``AES" to ``DES". Since ``DES" is a less secure algorithm, this can lead to vulnerabilities in the encryption implementation realized by this code. This does not represent a proper way to specify or select cryptographic algorithms, as it can easily be manipulated or misunderstood. \\
\midrule
MASC & String Case Transform & Cipher c = Cipher.getInstance(``des".toUpperCase(Locale.ENGLISH)); & TRUE & Gemini & FALSE & This code snippet does not contain any specific cryptographic misuses. & N/A \\

\bottomrule
\end{tabularx}
\end{sidewaystable}

\clearpage
\begin{sidewaystable}[ht]
\footnotesize
\centering
\vspace{8cm}
\begin{tabularx}{\textheight}{ >{\raggedright\arraybackslash}p{1cm} >
{\raggedright\arraybackslash}p{2.8cm} >
{\raggedright\arraybackslash}p{5.5cm} >
{\raggedright\arraybackslash}p{0.7cm} >
{\raggedright\arraybackslash}p{0.5cm} >
{\raggedright\arraybackslash}p{1cm} >
{\RaggedRight\arraybackslash}p{3cm} >
{\RaggedRight\arraybackslash}p{5.4cm}}
\toprule
MASC & Static IV & byte[] bytes = ``Hello".getBytes(); & TRUE & GPT & YES & Use of a static IV (Initialization Vector) & The code creates an \texttt{IvParameterSpec} using a static byte array derived from the string ``Hello". Using a static IV is a significant cryptographic misuse because IVs are supposed to introduce randomness to ensure that the same plaintext encrypted multiple times will yield different ciphertexts. This can lead to vulnerabilities such as revealing patterns in the plaintext and makes the encryption susceptible to certain attacks. A random or unique IV should be generated each time encryption is performed. \\
\midrule
MASC & Replace SHA256 to DES & cryptoDigest = MessageDigest.getInstance(``SHA-256".replace(``SHA-256", ``MD5")); & TRUE & Gemini & YES & Using a String manipulation technique to alter the algorithm name is a misuse. & The code attempts to dynamically change the MessageDigest algorithm from SHA-256 to MD5 using string manipulation. This is a cryptographic misuse because:Insecure Algorithm Selection: Hardcoding the algorithm name within the code makes it difficult to manage and update. It also increases the risk of accidentally using a weaker algorithm (MD5 in this case) compared to the intended stronger one (SHA-256). \\
\midrule
OWASP & BenchmarkTest00070.java & MessageDigest md = MessageDigest.getInstance(``SHA1", ``SUN"); & TRUE & GPT & YES & The code uses SHA1 for hashing sensitive data & SHA1 is considered cryptographically broken and unsuitable for further use. It is vulnerable to collision attacks, where two different inputs can result in the same hash output. In situations where security is critical (like storing hashes of passwords or sensitive information), a more robust hashing function such as SHA-256 or better yet, a key derivation function like PBKDF2, bcrypt, or Argon2 should be employed instead. \\
\midrule
OWASP & BenchmarkTest02291.java & Cipher c = Cipher.getInstance(``DES/CBC/PKCS5Padding", ``SunJCE"); & TRUE & Gemini & YES & Weak Encryption Algorithm and Key Size & The code uses the DES/CBC/PKCS5Padding algorithm for encryption. DES is a legacy algorithm with a small key size (56 bits) and is considered insecure by modern standards. It's vulnerable to brute-force attacks. Additionally, the key is generated using KeyGenerator.getInstance("DES") which reinforces the use of a weak key size. \\
\midrule
OWASP & BenchmarkTest00370.java & md = MessageDigest.getInstance(``sha-384", ``SUN"); & FALSE & GPT & YES & Using a weak hashing algorithm (SHA-384) to hash sensitive data. & While SHA-384 is a part of the SHA-2 family of cryptographic functions, using it alone for hashing passwords is considered a misuse because it is not a key derivation function and lacks sufficient resistance against brute-force attacks. Best practices recommend using algorithms specifically designed for hashing passwords, such as bcrypt, scrypt, or Argon2, which incorporate salting and an adjustable number of iterations to slow down attackers. \\
\midrule
OWASP & BenchmarkTest00270.java & MessageDigest md = MessageDigest.getInstance(``SHA-512", ``SUN"); & FALSE & Gemini & NO & N/A & There are no cryptographic misuses detected in the code. \\
\bottomrule
\end{tabularx}
\end{sidewaystable}

\end{document}